\documentclass{emulateapj}

\usepackage{graphicx}

\shorttitle{Force-free MHD Turbulence}
\shortauthors{Cho}

\begin{document}

\title{Simulation of Relativistic Force-free Magnetohydrodynamic 
Turbulence}

\author{Jungyeon Cho}
\affil{CITA, Univ. of Toronto, 60 St. George St., Toronto,
              ON M5S 3H8, Canada; cho@cita.utoronto.ca}

\begin{abstract}
We present numerical studies of 3-dimensional
magnetohydrodynamic (MHD) turbulence in a strongly magnetized medium
in the extremely relativistic limit, in which
the inertia of the charge carriers can be neglected.
We have focused on strong Alfvenic turbulence in the limit.
We have found the following results.
First, the energy spectrum is consistent with a Kolmogorov spectrum:
$E(k)\sim k^{-5/3}$.
Second, turbulence shows a Goldreich-Sridhar type
anisotropy: $k_{\|} \propto k_{\perp}^{2/3}$, where
$k_{\|}$ and $k_{\perp}$ are wavenumbers along and perpendicular to
the local mean magnetic field directions, respectively.
These scalings are in agreement with earlier theoretical predictions
by Thompson \& Blaes.
\end{abstract}

\keywords{MHD --- turbulence ---  relativity}
\section{Introduction}
In a strongly magnetized extremely relativistic medium, as in
the magnetospheres of pulsars and black holes,
energy density of the conducting matter can be neglected.
When this is the case, 
the Lorentz force vanishes 
and, thus, we can use force-free approximation for the medium.
Turbulence in this force-free magnetohydrodynamics (MHD) limit is the
topic of this paper.
The magnetospheres of pulsars and black holes 
(see Goldreich \& Julian 1969;
Blandford \& Znajek 1977; Duncan \& Thompson 1992) 
or gamma-ray bursts (see Thompson 1994;
Lyutikov \& Blandford 2003) 
are examples of relevant astrophysical systems.

The force-free MHD admits two normal modes, Alfven and fast modes
(see, for example, Thompson \& Blaes 1998; Komissarov 2002).
In the presence of a strong 
mean magnetic field ${\bf B}_0=B_0 \hat{\bf z}$,
nonlinear interactions of these modes result in turbulence cascade.
In this paper, we focus on Alfvenic turbulence.

In the non-relativistic limit, Goldreich \& Sridhar (1995) model
provides a good theoretical description of strong Alfvenic turbulence.
The model predicts a Kolmogorov energy spectrum ($E(k)\propto k^{-5/3}$)
and a scale-dependent anisotropy ($k_{\|} \propto k_{\perp}^{2/3}$).
Here
$k_{\|}$ and $k_{\perp}$ are wavenumbers along and perpendicular to
the local mean magnetic field directions, respectively.
The scale-dependent anisotropy states that anisotropy is
more pronounced at smaller scales.
The model was first numerically confirmed by Cho \& Vishniac (2000).
Subsequent numerical studies and further discussions are given in 
Maron \& Goldreich (2001),
Cho, Lazarian, Vishniac (2002), Cho, Lazarian, Vishniac (2003).

Thompson \& Blaes (1998) theoretically studied physics of 
force-free MHD turbulence.
They introduced two new formulations for the force-free MHD and
discussed various aspects of turbulence processes, including
those of Alfvenic turbulence.
They found that force-free Alfvenic MHD turbulence exhibits
scaling relations very similar to the non-relativistic ones.
That is to say, regarding to
strong force-free Alfvenic MHD turbulence,
they argued that
it has
a Kolmogorov spectrum and a Goldreich-Sridhar type anisotropy.

In this paper, we numerically study strong force-free Alfvenic turbulence
in flat space.
In \S2, we describe our numerical methods.
In \S3, we present results.
We give discussion in \S4 and conclusion in \S5.

\section{Method}
We solve the following system of equations:
\begin{equation}
   \frac{ \partial {\bf Q} }{ \partial t }
 + \frac{ \partial {\bf F} }{ \partial x^1 }
 =0,
\end{equation}
where
\begin{eqnarray}
{\bf Q}=(S_1,S_2,S_3,B_2,B_3), \\
 {\bf F}=(T_{11},T_{12},T_{13},-E_3,E_2), \\
 T_{ij}=-(E_iE_j + B_iB_j)+\frac{ \delta_{ij} }{2} (E^2+B^2), \\
 {\bf S}={\bf E}\times {\bf B}, \\
 {\bf E}=-\frac{1}{B^2} {\bf S}\times {\bf B},
\end{eqnarray}
where ${\bf E}$ is the electric field and 
${\bf S}$ the Poynting flux vector
(see Komissarov 2002).
One can derive this system of equations from
\begin{eqnarray}
 \partial_{\mu} {}^{*}F^{\mu\nu}=0 \mbox{~~~(Maxwell's eq.)}, \\
 \partial_{\mu} F^{\mu\nu}=-J^{\nu} \mbox{~~~(Maxwell's eq.)}, \\
 \partial_{\mu} T^{\nu\mu}_{(f)}=0 \mbox{~~~(energy-momentum eq.)}, \\
                F_{\nu\mu} u^{\mu}=0 \mbox{~~~(perfect conductivity)},
    \label{eq_cond}
\end{eqnarray}
where ${}^{*}F^{\mu\nu}$ is the dual tensor of the electromagnetic field,
$u^{\mu}$ the fluid four velocity, and $T^{\mu\nu}_{(f)}$ 
the stress-energy tensor of the electromagnetic field:
\begin{equation}
 T^{\mu\nu}_{(f)}=F^{\mu}_{\alpha}F^{\alpha \nu} -\frac{1}{4}
                 (F_{\alpha \beta}F^{\alpha \beta})g^{\mu\nu},
\end{equation}
where $g^{\mu\nu}$ is the metric tensor and $F^{\alpha\beta}$ is
the electromagnetic field tensor.
We ignore the stress-energy tensor of matter.
We use flat geometry and units with $c=1$.
Greek indices run from 1 to 4.
One can obtain the force-free condition from 
Maxwell's equations and the energy-momentum equation:
$\partial_{\mu} T^{\nu\mu}_{(f)}=-F_{\nu\mu}J^{\mu}=0$.
{}From equation(\ref{eq_cond}), one can derive
\begin{eqnarray}
 {\bf E}\cdot {\bf B}=0, \\
 B^2-E^2 > 0.
\end{eqnarray}

We solve equations (1)-(6) using a MUSCL-type scheme with HLL fluxes
(Harten et al. 1983; in fact in force-free MHD these fluxes reduce to
Lax-Friedrichs fluxes)
and monotonized central limiter (see Kurganov et al. 2001). 
The overall scheme is second-order accurate.
After updating the system of equations along $x^1$ direction,
we repeat similar procedures for $x^2$ and $x^3$ directions with
appropriate rotation of indexes.
Gammie, McKinney, \& T\'oth (2003) used a similar scheme for general relativistic MHD and Del Zanna, Bucciantini, \& Londrillo (2003) used
a similar scheme to construct a higher-order scheme for special
relativistic MHD.

While the magnetic field consists of the uniform background field and a
fluctuating field, ${\bf B}= {\bf B}_0 + {\bf b}$, electric field
has only fluctuating one.
The strength of
the uniform background field, $B_0$, is set to 1.
We use a grid of $512^3$.
At $t=0$, only Alfven modes are present in the range
\begin{eqnarray}
  4\leq k_{\perp} \leq 6 \mbox{~~~and} \\
  1\leq k_{\|} \leq 2
\end{eqnarray}
in wavevector ({\bf k}) space.
The MHD condition ${\bf E} \cdot {\bf B}=0$ is satisfied at $t=0$.
The energy density of the random magnetic and electric fields at $t=0$ 
is $\sim 0.1$. Therefore, we have
\begin{equation}
   \chi \equiv \frac{ b k_{\perp} }{ B_0k_{\|} } \sim 1
\end{equation}
at $t=0$.

\section{Results for Strong Turbulence}
Figure 1 shows energy spectra of magnetic field.
At $t=0$ (not shown) 
only large scale (i.e. small $k$)
Fourier modes are excited. 
At later times, energy cascades down to small scale 
(i.e. large $k$) modes.
After $t \sim 3$,
the energy spectrum decreases without changing its slope.
The spectrum at this stage is very close to a Kolmogorov spectrum:
\begin{equation}
  E(k) \propto k^{-5/3}.
\end{equation}

In Figure 2, we plot contour diagram of the second-order
structure function for magnetic field
in a local frame, which is aligned with the local mean magnetic field
${\bf B_L}$:
\begin{equation}
    \mbox{SF}_2(r_{\|},r_{\perp})=<|{\bf B}({\bf x}+{\bf r}) -
                 {\bf B}({\bf x})|^2>_{avg.~over~{\bf x}},
\end{equation}
where ${\bf r}=r_{\|} {\hat {\bf r}}_{\|} +r_{\perp} {\hat {\bf r}}_{\perp}$
and ${\hat {\bf r}}_{\|}$ and ${\hat {\bf r}}_{\perp}$ are unit vectors
parallel and perpendicular to the local mean field ${\bf B_L}$, respectively.
See Cho et al. (2002) 
and Cho \& Vishniac (2000) for the 
detailed discussion of the local frame.
          The contour plot clearly
          shows existence of scale-dependent anisotropy:
          smaller eddies are more elongated.
By analyzing the relation between the semi-major axis 
($\sim l_{\|} \sim 1/k_{\|}$) and the semi-minor
axis ($\sim l_{\perp} \sim 1/k_{\perp}$) of the contours, 
we can obtain the 
relation between $k_{\|}$ and $k_{\perp}$.
The result in Figure 3 is consistent with the Goldreich-Sridhar type
anisotropy:
\begin{equation}
   k_{\|} \propto k_{\perp}^{2/3}.
\end{equation}

The electric field shows similar scalings as the magnetic field.
Our numerical results for spectrum and anisotropy are
consistent with theoretical predictions by Thompson \& Blaes (1998).

\begin{figure*}[t]
\plottwo{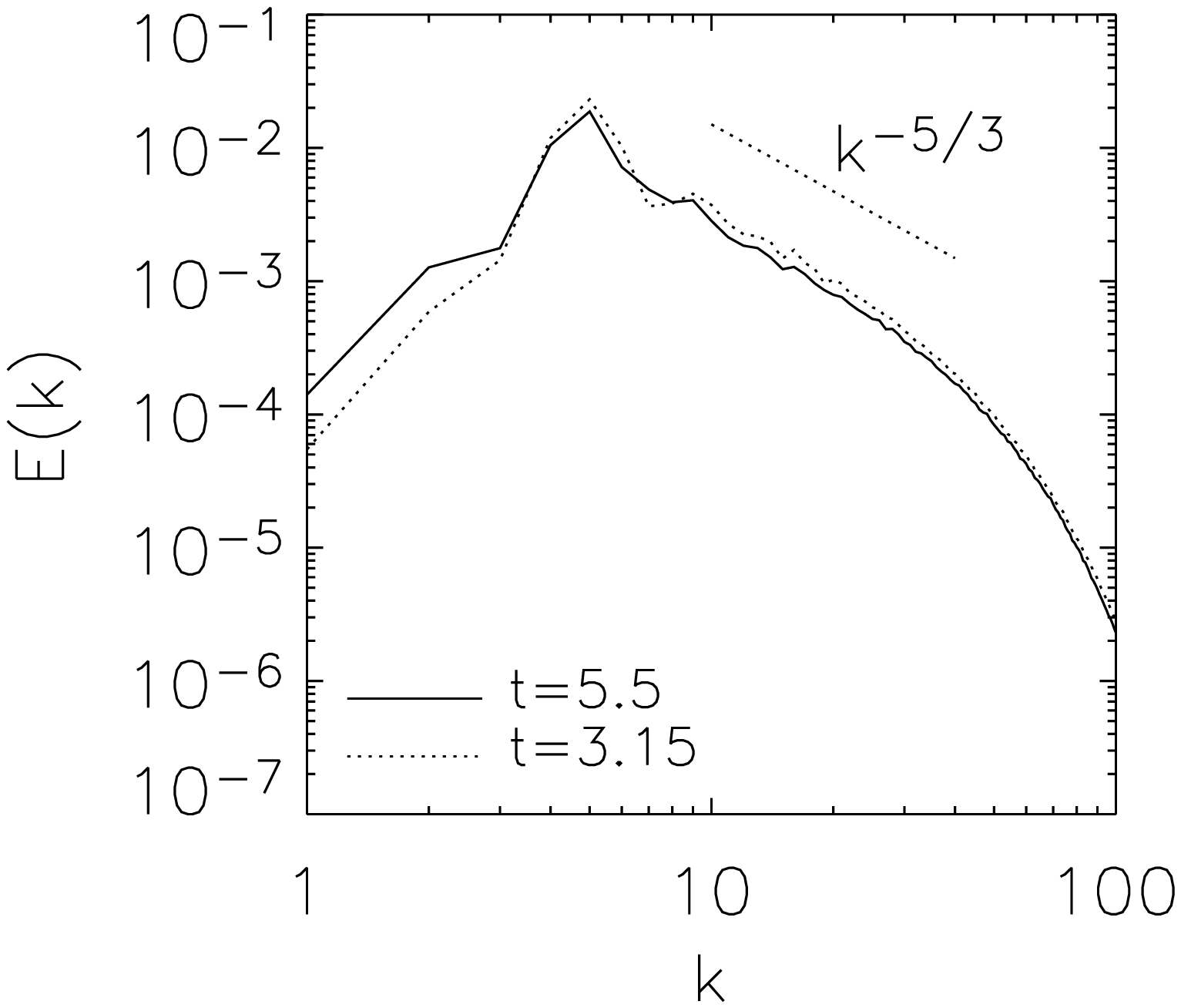}{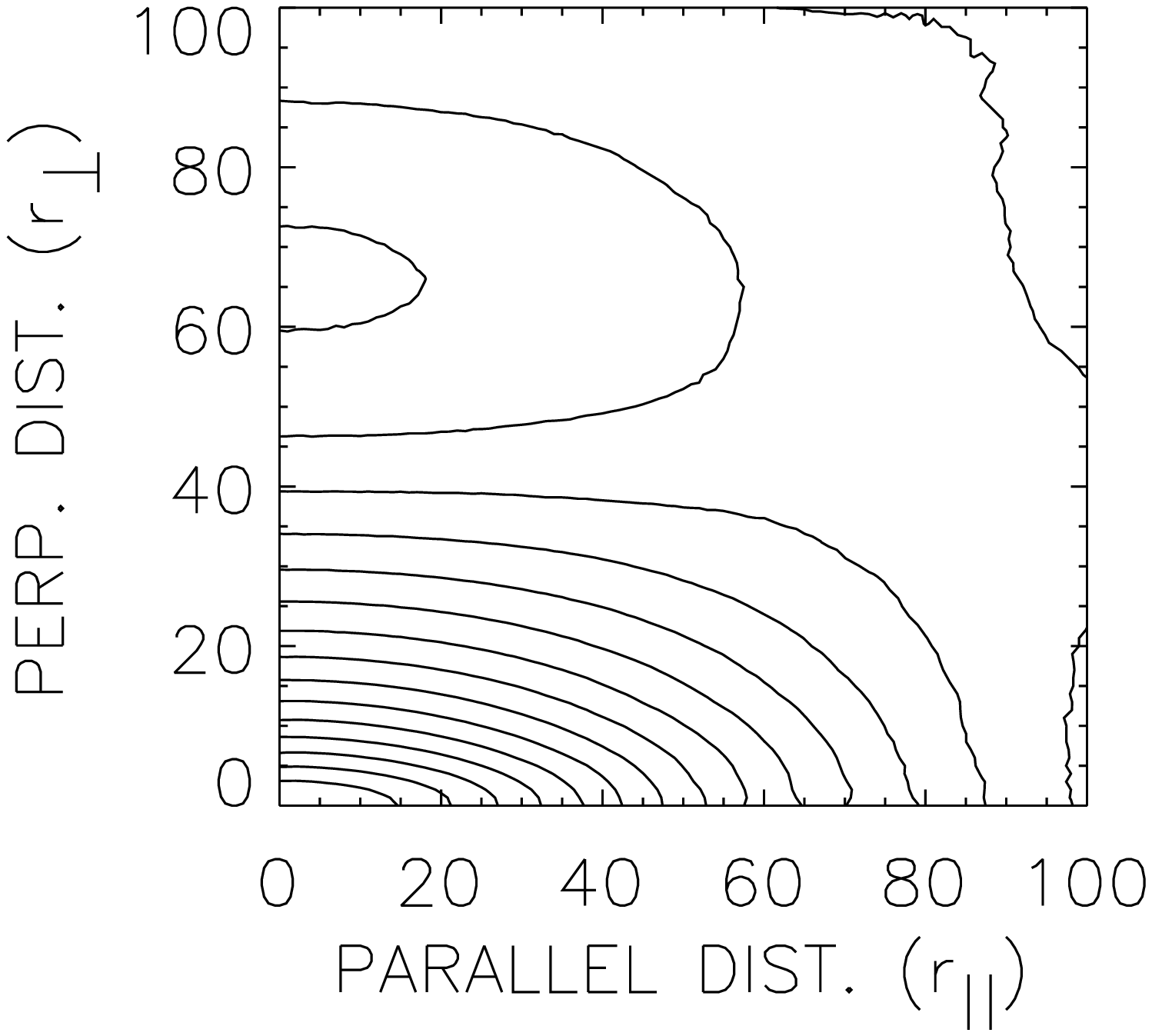}
\caption{ 
         Spectra at t=3.15 (dotted) and t=5.5 (solid). 
         The spectra are compatible with a Kolmogorov
         spectrum: $E(k)\propto k^{-5/3}$.
}
\caption{
         Contour plot of the second structure function for the 
         magnetic field at $t=4.7$. Contours, representing eddy shapes,
         clearly show scale-dependent anisotropy: smaller contours are
         more elongated.
}
\end{figure*}

\section{Discussion}

\subsection{Phenomenology}
Thompson \& Blaes (1998) derived scaling relations for force-free MHD
turbulence, which are confirmed by our simulations.
Here we re-derive the scaling relations using a simple phenomenology.

The magnetic and the electric fields follow the equations:
\begin{eqnarray}
 \frac{ \partial{\bf B} }{ \partial t }
  = -\nabla \times {\bf E}, \label{eq_dbdt} \\
 \frac{ \partial{\bf E} }{ \partial t }
  = \nabla \times {\bf B} - {\bf J},  \label{eq_dedt}
\end{eqnarray}
where the current density ${\bf J}$ is 
\begin{equation}
  {\bf J} = \frac{ ({\bf E}\times {\bf B})\nabla \cdot {\bf E} +
                   ( {\bf B}\cdot \nabla \times {\bf B}
                    -{\bf E}\cdot \nabla \times {\bf E}) {\bf B} }
                  {B^2}
   \label{eq_j}
\end{equation}
(Lyutikov 2003).
The current density ${\bf J}$ in equation (\ref{eq_dedt}) is essential
for nonlinear interactions.
In equation (\ref{eq_j}), when $B_0 \gg b$ and $k_{\perp} \gg k_{\|}$,
the first term on the right side dominates for Alfven turbulence.

Suppose that we have a Alfven wave packet whose parallel size is 
$l_{\|}\sim k_{\|}^{-1}$ and
perpendicular size $l_{\perp}\sim k_{\perp}^{-1}$ ($\sim l\sim k^{-1}$ when
anisotropy is present).
This wave packet travels along the magnetic field line at the speed of
$c~(=1)$.
When this wave packet collides with opposite-traveling wave packets (of
similar size), the change of energy ($\Delta \mathcal{E}$) per collision is
\begin{equation}
  \Delta \mathcal{E}  \sim ({d\mathcal{E}}/{dt}) \Delta t  \sim 
(k b_l^3/B_0)/k_{\|},  \label{eq_Den}
\end{equation}
where we use equation (\ref{eq_dedt}) to estimate
${d\mathcal{E}}/{dt}$. 
We assume  $B_0 \gg b$ and $k_{\perp} \gg k_{\|}$, and, hence,
$J \sim kb^2/B_0$
(see equation (\ref{eq_j})).
Note that $b_l \sim E_l$.
Therefore, 
\begin{equation}
   \Delta \mathcal{E}/\mathcal{E} \sim (k b_l)/(k_{\|}B_0) \sim t_w/t_{eddy},  \label{eq_dee}
\end{equation}
where $t_w ~(\equiv l_{\|}/c=1/k_{\|})$ is the wave period and
$t_{eddy} ~(\equiv l_{\perp}/v_l \sim 1/(k_{\perp} b/B_0))$ the eddy
turnover time. Here we used $v_l \sim cb/B_0 = b/B_0$.
We consider two cases: $\chi \equiv (k b_l)/(k_{\|}B_0) \sim 1$
and $\chi < 1$.

If $\chi ~(\equiv  k b_l/k_{\|}B_0)  \sim 1$, the resulting turbulence 
is strong (see Goldreich \& Sridhar 1995 for more discussion
about {\it non-relativistic} Alfvenic turbulence and the role of $\chi$).
Our simulation satisfies this condition (Figure 4).
Since $\chi \sim t_w/t_{eddy}$ (equation [\ref{eq_dee}]),
the condition $\chi \sim 1$ implies that there is a balance between
the hydrodynamic and the wave times-scales, which means that
we can use either time-scale for the energy cascade time-scale.
Therefore, the constancy of spectral energy cascade rate in this case
becomes
\begin{equation}
 \frac{ b_l^2 }{ t_{cas} } \sim b_l^2 v_l k \propto b_l^3 k =  
 \mbox{constant},
\end{equation}
which results in
\begin{equation}
  b_l \propto l^{1/3} \mbox{~~or~~} E(k) \propto k^{-5/3},
\end{equation}
where we use $kE(k)\approx b_l^2$.
{}From $k b_l/(k_{\|}B_0) \sim 1$, we can obtain
\begin{equation}
 k_{\|} \propto k_{\perp}^{2/3},
\end{equation}
where we use $k\sim k_{\perp}$.

If $\chi < 1$, then the turbulence is weak: many collisions are required
to make $\Delta \mathcal{E} /\mathcal{E} \sim 1$.
Thompson \& Blaes (1998) argued that force-free MHD turbulence
in this regime has virtually constant $k_{\|}$ and $E(k)\propto k^{-2}$.
See Galtier et al. (2000) for {\it non-relativistic} weak MHD 
turbulence.

\begin{figure*}[t]
\plottwo{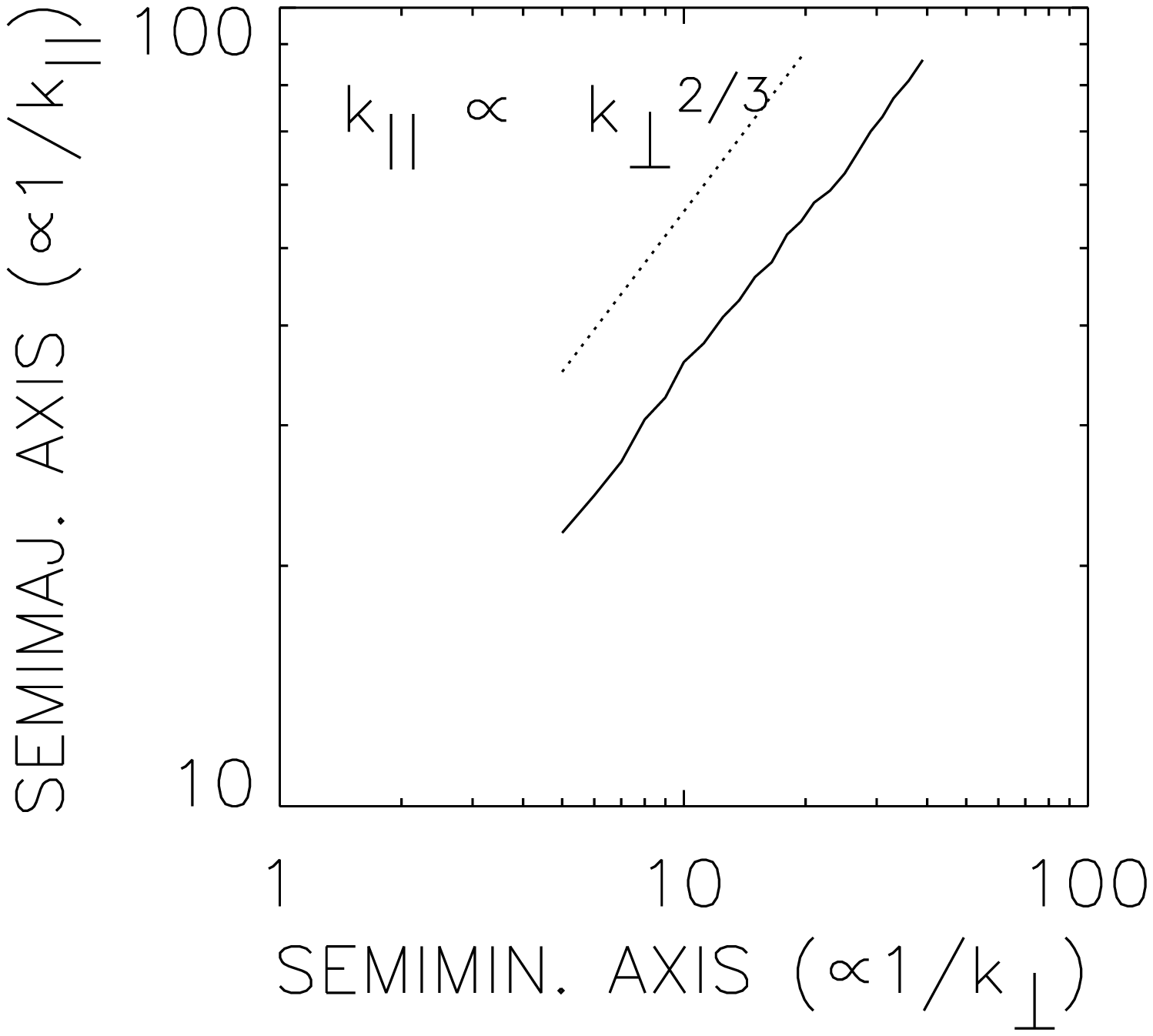}{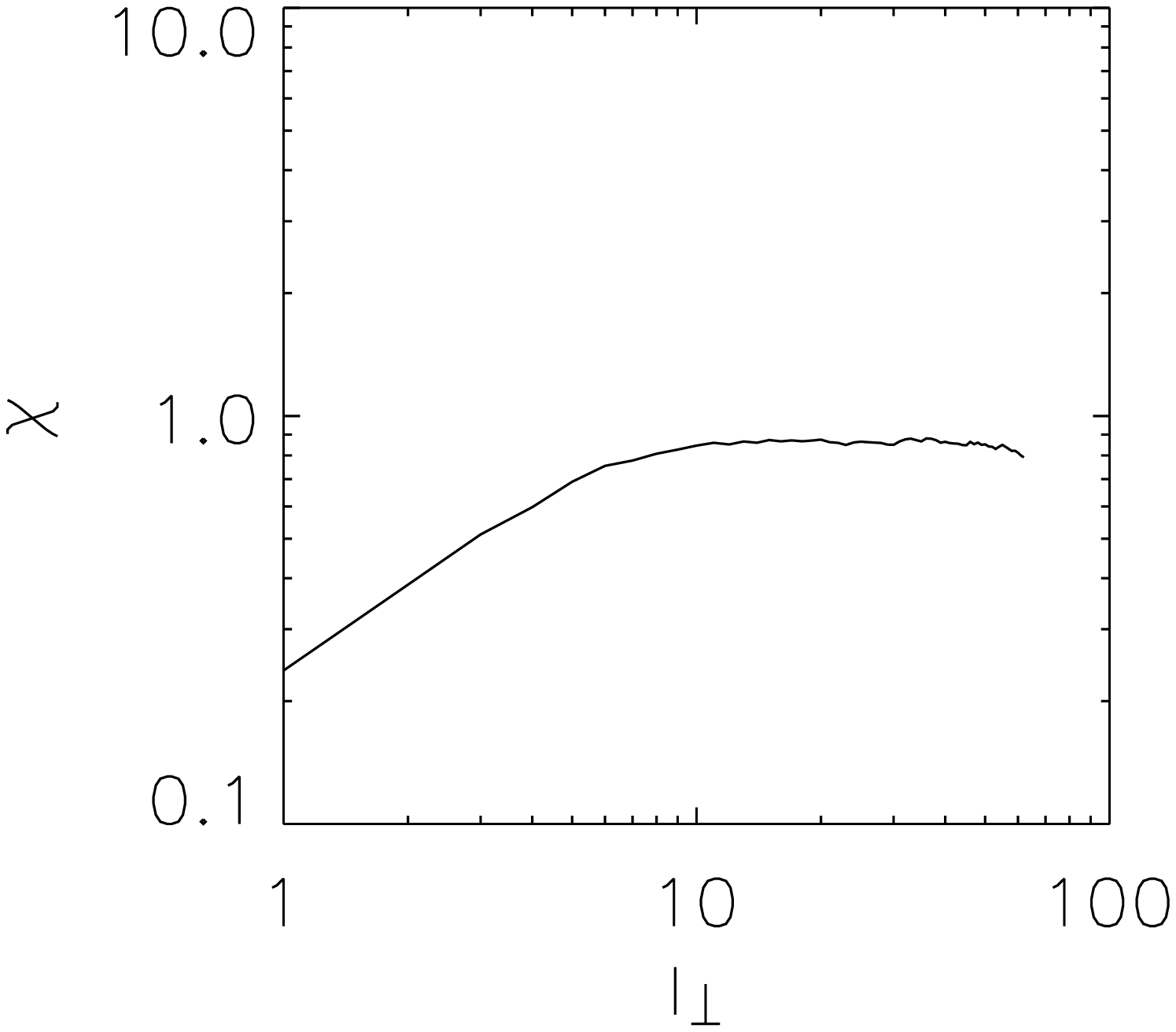}
\caption{
         Anisotropy. Semi-major and semi-minor axes are obtained from
         Figure 2. We observe a Goldreich-Sridhar type anisotropy:
         $k_{\|}\propto k_{\perp}^{2/3}$.
}
\caption{
         $\chi$ $(\equiv  k b_l/k_{\|}B_0 \sim 
         [l_{\|}/l_{\perp}][SF_2(0,l_{\|})]^{1/2})$ is nearly constant
        in our simulations.  
        We take the ratio $l_{\|}/l_{\perp}$ from contours in
        Figure 2. Note that the x-axis is $l_{\perp} (\sim 1/k_{\perp})$ 
        in grid units.
}
\end{figure*}

\begin{figure*}
\plottwo{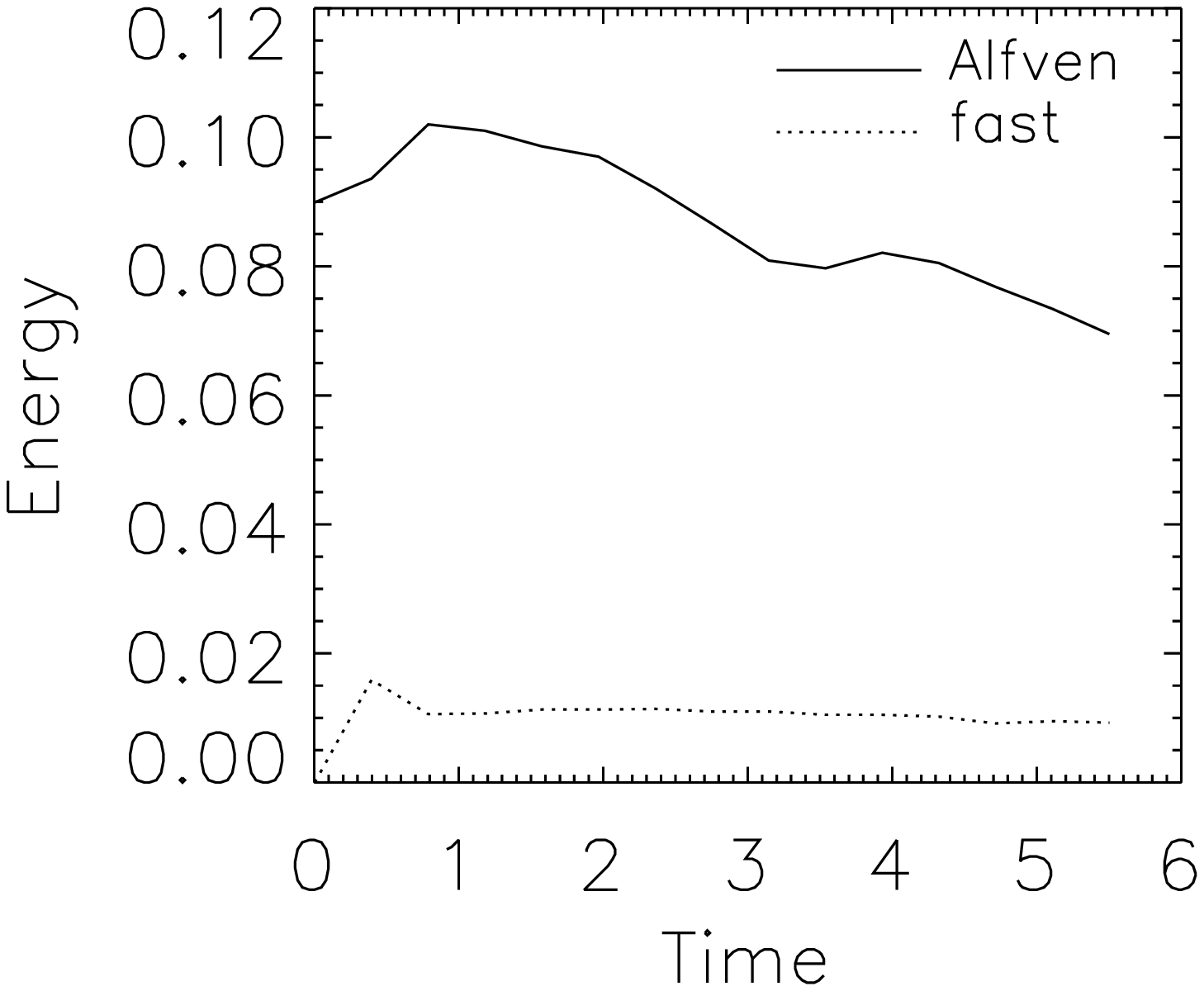}{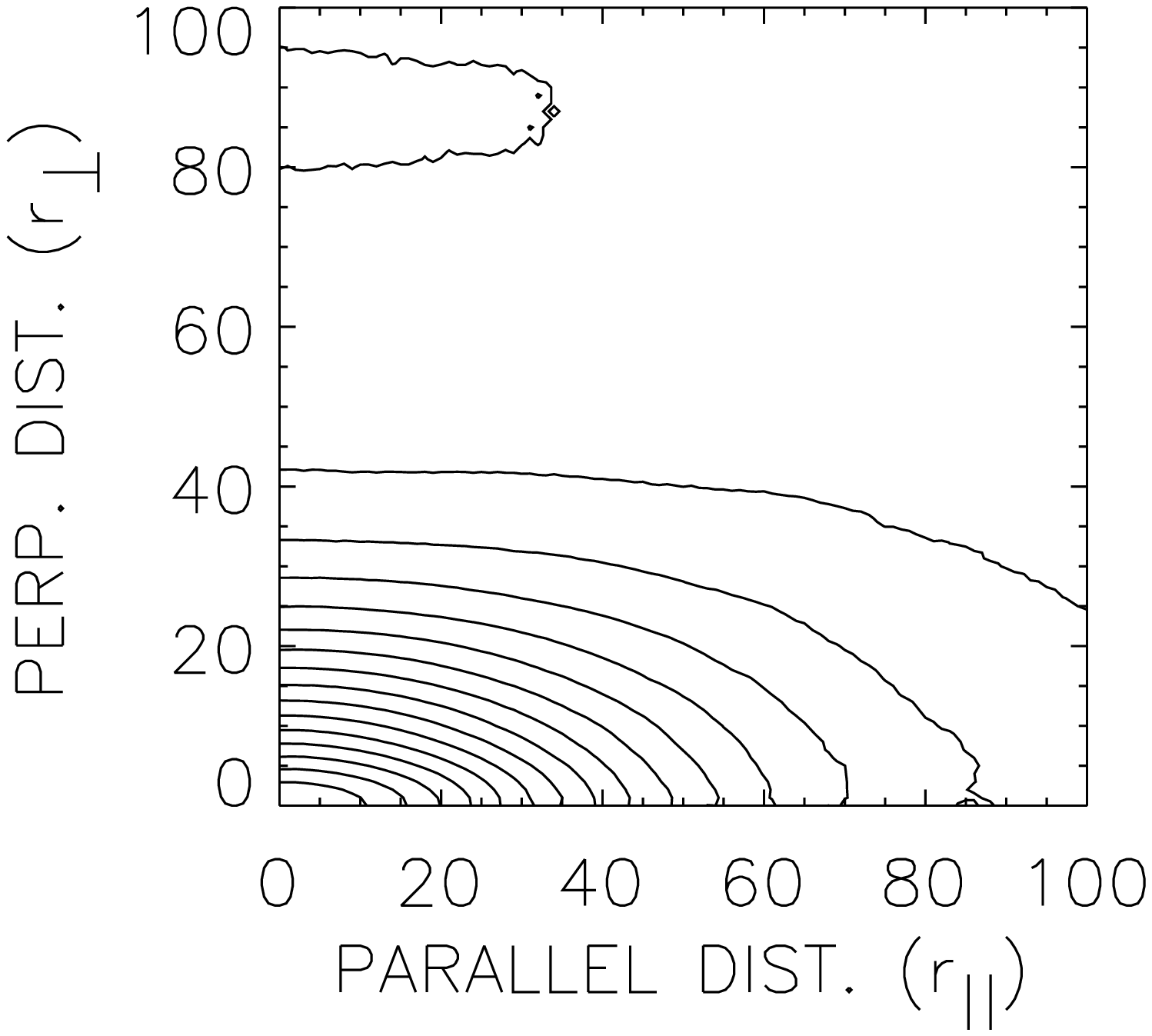}
\caption{ 
         Generation of fast modes. 
         Energy in fast modes grows rapidly initially.
         After $t \gtrsim 0.5$, the energy ratio of
         fast to Alfven modes is about $\sim 0.13$.
}
\caption{
         Contour diagram of the second-order structure function
         for fast modes at $t=4.7$. Contours show anisotropy that is
         virtually scale-independent.
}
\end{figure*}

\subsection{Scaling of fast modes}
At the beginning of the simulation, we have only Alfven modes.
As later times nonlinear interactions of Alfven wave packets 
produce fast modes. Figure 5 shows the amount of fast modes
generated from the Alfvenic turbulence.
The ratio of fast to Alfven energy is roughly 0.13 to 0.15, which
can be a measure of mode coupling between Alfven and fast modes.
See Thompson \& Blaes (1998) for detailed discussion of
mode coupling between Alfven and fast modes.
It is interesting that the energy ratio is not very much different from
the energy ratio of compressible to incompressible modes in
hydrodynamic (Porter, Woodward, \& Pouquet 1998), 
super-Alfvenic non-relativistic MHD (Boldyrev, Nordlund, Padoan, 2002), or
mildly sub-Alfvenic non-relativistic MHD (Cho \& Lazarian 2002) cases.
Unlike our current simulation, all these works used either isotropic
initial condition or isotropic driving.

Cho \& Lazarian (2002) showed that fast modes generated 
from solenoidal components
in {\it non-relativistic} MHD turbulence
are isotopic when driving or initial condition is isotropic.
In case of anisotropic driving or an anisotropic initial condition,
fast modes can be anisotropic. But we expect that anisotropy of
fast modes in that case is scale-{\it independent}.

Can we derive a similar conclusion for force-free MHD?
In Figure 6, we plot contour diagram of the second-order
structure function for fast modes.
As in the case of Alfven modes, we can assume that contours
represent eddy shapes.
Contours do show anisotropy. 
But, the anisotropy does not show strong scale-dependence.

\subsection{scaling of current}
In equation (\ref{eq_dedt}), we are not interested in the parallel
components of the right hand side of the equation. 
Perpendicular components are important for the evolution of the electric 
field. Therefore let us consider only the perpendicular components
of the equation. We also assume that we have only Alfven modes.
Then the three terms on the right hand side of equation (\ref{eq_j})
can be approximated by
\begin{eqnarray}
  ({\bf E}\times {\bf B})\nabla \cdot {\bf E}/B^2 \sim kb_l^2/B_0, 
   \label{eq_j1st} \\
  ({\bf B}\cdot \nabla \times {\bf B}){\bf B}/B^2 \sim k_{\|}b_l^2/B_0, 
   \label{eq_j2nd} \\
  ({\bf E}\cdot \nabla \times {\bf E}){\bf B}/B^2 \sim k b_l^3/B_0^2,
\end{eqnarray}
where we assumed $b_l\sim E_l$.
Therefore, the dominant term is the first term.
{}From this we can easily show that the spectrum of the current density
follows
\begin{equation}
 J_l \propto l^{-1/3}, \mbox{~~~ or ~~} E_{J}(k)\propto k^{-1/3}.
\end{equation}
On the other hand, the charge density follows
\begin{equation}
 \rho_e \propto \nabla \cdot {\bf E} \propto k^{2/3},
\end{equation}
or 
\begin{equation}
  E_{\rho,e}(k) \propto k^{1/3}.
\end{equation}

\section{Conclusion}
Using numerical simulations, we have studied 3 dimensional force-free MHD turbulence.
 We have found that
energy spectrum is compatible with a Kolmogorov spectrum: 
$E(k)\propto k^{-5/3}$.
We have calculated anisotropy and found a Goldreich-Sridhar type
anisotropy: $ k_{\|}\propto k_{\perp}^{2/3}$.
These findings are consistent with earlier theoretical studies in
Thompson \& Blaes (1998).

\begin{acknowledgments}  
I thank Chris Thompson for useful
discussions and advices. I also thank Maxim Lyutikov 
for useful discussions.
This work utilized CITA supercomputing facilities.
\end{acknowledgments}

\end{document}